\documentclass{PoS}

\usepackage{graphics,graphicx}
\usepackage{epsfig}
\usepackage{mathptmx}

\title{Nuclear lines revealing the injection of cosmic rays in supernova remnants}

\ShortTitle{Nuclear lines revealing the injection of cosmic rays in supernova remnants}

\author{Omar Tibolla, Karl Mannheim, Alexander Summa, Aleksander Paravac\\
        Institut f\"ur Theoretische Physik und Astrophysik, Universit\"at W\"urzburg, Germany \\
        E-mail: \email{omar.tibolla@gmail.com ; Omar.Tibolla@astro.uni-wuerzburg.de }}

\author{Jochen Greiner, Gottfried Kanbach\\
        Max Plank Institut f\"ur Extraterrestrische Physik, Garching, Germany}

\abstract{At high energies, the hadronic origin of gamma rays from supernova remnants is still debated. Assuming the observed gamma-rays from the Wolf-Rayet supernova remnant Cas A are due to accelerated protons and ions, we predict the nuclear de-excitation line emission arising from interactions with the heavy elements in the supernova ejecta. \\
This illustrative example highlights the importance of MeV gamma ray observations of the hadronic fingerprint of cosmic rays. The lines could be observed in the MeV band with a future space mission such as GRIPS which has been proposed to ESA as an all-sky survey mission with gamma-ray, X-ray and near-infrared telescopes. In MeV gamma rays, its sensitivity will improve by a factor of 40 compared with previous missions.}

\FullConference{25th Texas Symposium on Relativistic Astrophysics - TEXAS 2010\\
		December 06-10, 2010\\
		Heidelberg, Germany}

\begin{document}

\section{GRIPS}

The Gamma-Ray Imaging, Polarization and Spectroscopy (GRIPS) mission \cite{1} consists of a Gamma-ray monitor (GRM), an X-ray monitor (XRM) and an Infrared telescope (IRT):
\begin{itemize}
\item {the IRT is an IR telescope with a 1 meter mirror and it will be a copy of Euclid \cite{2}; in addition, in the IRT focal plane there is set a GROND \cite{3}-like system with 7 detectors, with a 10$' \times $10$'$ field of view;}
\item{the XRM is re-using eRosita \cite{4} technology; it will work in the energy range 0.1 - 10 keV with a spatial resolution of $\sim $30$''$;}
\item{the GRM will work in the energy range 200 keV - 80 MeV (the energy range upper limit is easily extendable to higher energies) with a spatial resolution of $\sim $1$^{\circ}$ and it will also provide polarization ($\sim$ 1\%) measurement capability.}
\end{itemize}

\begin{figure}
\begin{center}
\hbox{
\psfig{figure=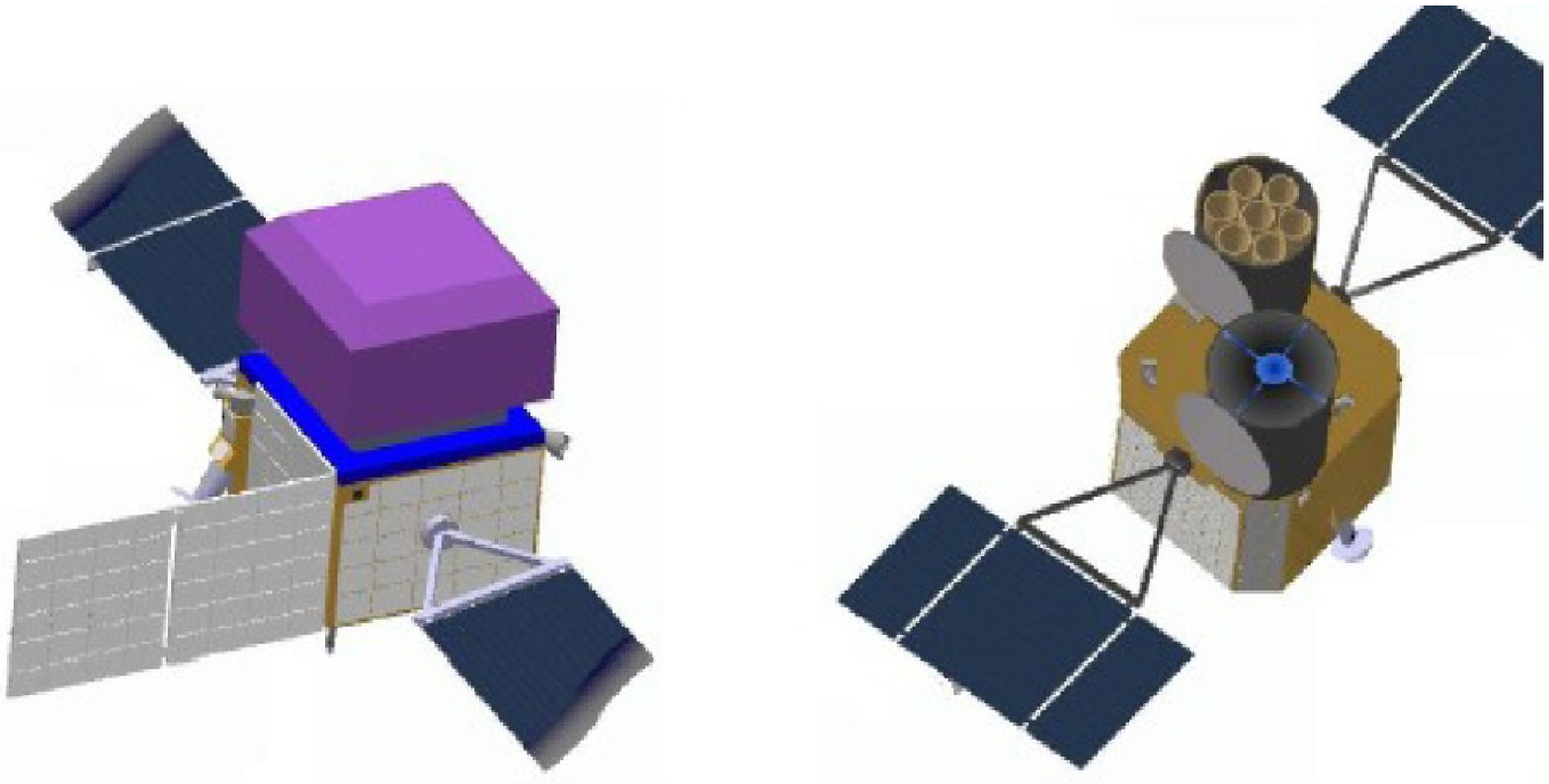,height=3.5cm,angle=0}
\hspace{1.5cm}
\psfig{figure=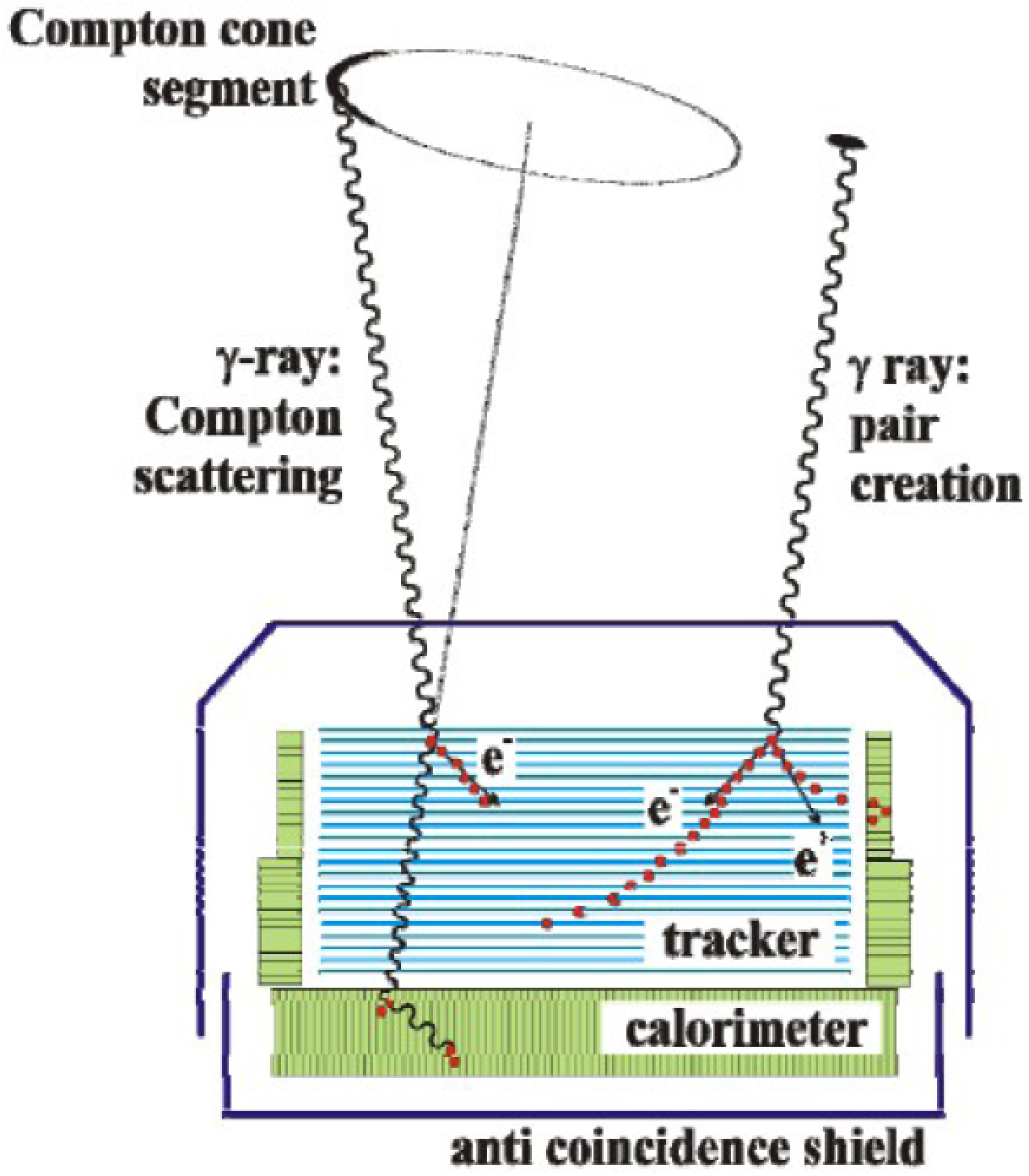,height=6.0cm, angle=0}
\vspace{-1cm}
}
\end{center}
\caption{\footnotesize
{{\it Left panel}: Schematic representation of the two satellite busses.
{\it Right panel}: Detail on GRM working principle and its schematic representation.
}
}
\label{fig1}
\end{figure}

These 3 instruments will be set in orbit onboard 2 different satellite buses, however, respecting the Soyuz weight limits (the total weight of GRIPS is 5 tons, while the Soyuz- Fregat payload limit is 5.2 tons), it will require one single Soyuz launcher. Figure \ref{fig1} show a schematic representation of the two satellite buses, one carrying GRM ,while the other for IRT and XRT, and a cartoon zooming into the GRM working principle.

GRM is a combined Compton and pair conversion telescope, consisting of 3 different parts:
\begin{itemize}
\item {A Tracker (TKR), made of Silicon double-side detectors (DSD), where the initial interaction takes place and the secondary charged particles are tracked; the TKR consists of 4 towers, each tower consists of 64 layers spaced 5 mm from each other (also a configuration with 3 mm spacing has been under investigation) and each layer consists of 4$\times $4 DSD wafers of 10 $\times$10 cm$^2$ each.}
\item {A Calorimeter (CAL), made of La$^3$Br, where the energy of the secondary particles is converted into light signal.}
\item {An Anti-coincidence Shield (ACS), made of plastic scintillator (Ne110), that will shield the instrument from charged particles; note that the efficient background rejection, provided by the ACS, is a crucial point for an MeV telescope. GRIPS is a mature project that is planned to be launched in 2020.}
\end{itemize}

A detailed description of GRIPS mission can be found in \cite{5} and in the original project \cite{joc}.

\section{Origin of cosmic rays}

One of the most important achievement of GRIPS will regard cosmic rays (CRs). CRs are 98\% hadrons and 2\% leptons. This discovery traces back one century to the early works of Domenico Pacini \cite{6} and Victor Hess \cite{7} but their origin is still unknown. While the lepton acceleration has been proven (e.g. by X-rays and Radio observations of SNRs or PWNe), the ``smoking gun'' for the hadronic acceleration is still missing. Since the sixties, the SN explosions are thought to be responsible for the hadronic component of CRs \cite{8} and the recent discoveries at TeV (from Imaging Atmospheric Cherenkov Telescpes, IACTs) and GeV ({\it Fermi} LAT) energies seem to support this, however there is no conclusive proof yet. In the future the ``smoking gun'' could also come from the neutrino telescopes (such as Icecube and KM3NET) despite their limitations due to the small cross section of neturino interactions.
GRIPS will provide this proof in two ways:
\begin{itemize}
\item {by distinguishing between hadronic and leptonic emission from the continuum spectrum.}
\item {by studying the nuclear de-excitation lines.}
\end{itemize}

\section{Cosmic rays: continuum spectra}

\begin{figure}
\begin{center}

\psfig{figure=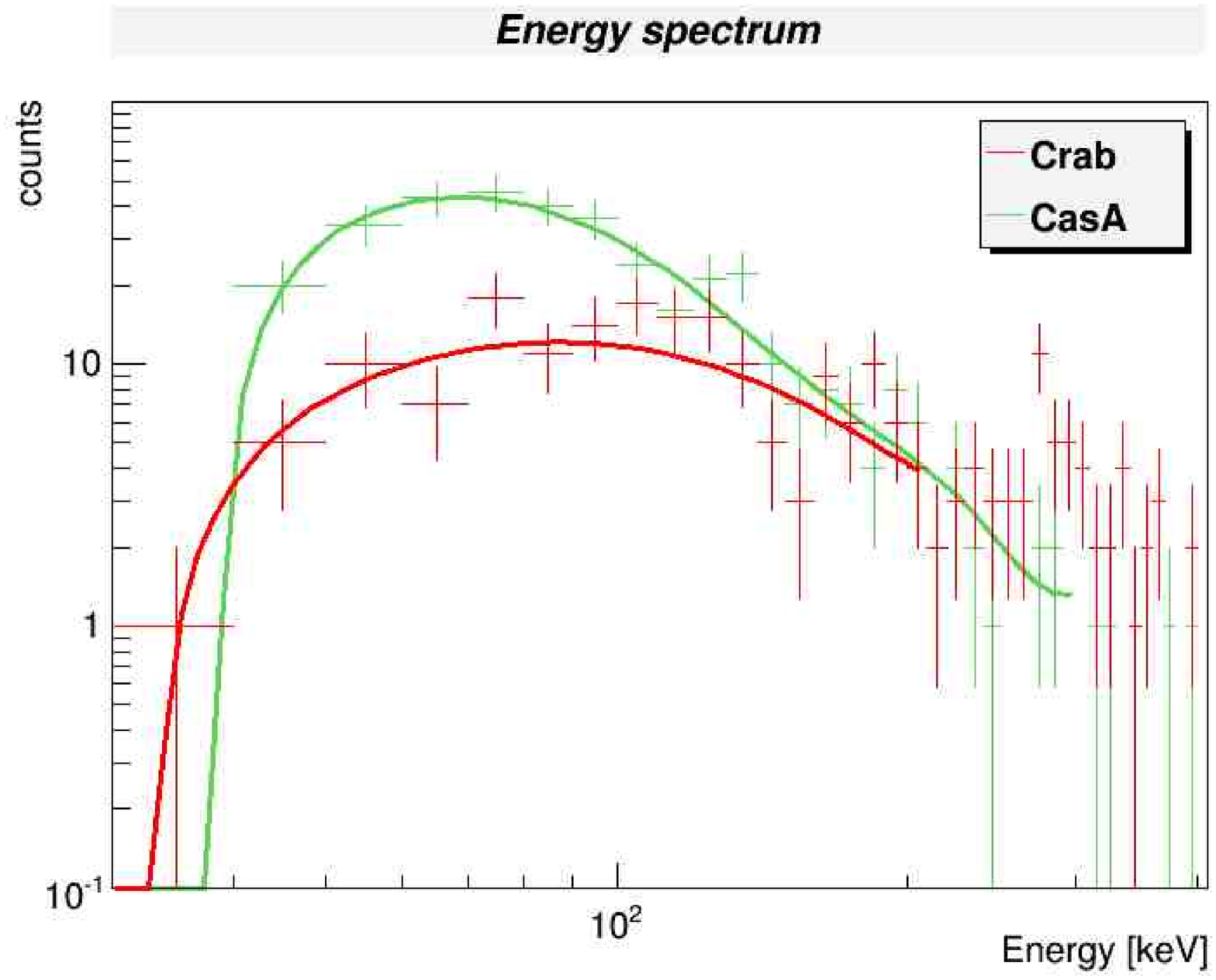,height=5.0cm,angle=0}
\vspace{-0.5cm}

\end{center}
\caption{\footnotesize
{Preliminary simulation of few minutes of {\it GRIPS} observations using MEGAlib software \cite{megalib}: Crab Nebula and Cassiopeia A spectra.The spectra are normalized in order to show similar energy fluxes for the sake of demonstration of the instrument capabilities to distinguish between Cas A - like spectra and Crab-like ones.
}
}
\label{cont}
\end{figure}

Leptonic and hadronic processes have different signatures in the GRM energy range. GRIPS will be able to detect the sources that are thought to be responsible of CRs acceleration and to precisely measure their spectra. Figure \ref{cont} shows the spectra of the Crab Nebula and Cassiopeia A extracted from the preliminary simulation of few minutes of GRIPS observations: the spectra are normalized in order to show similar energy fluxes for the sake of demonstration of the instrument response to a Cas A - like spectrum and its capabilities to distinguish it from a Crab-like spectrum. Let us underline once more the polarimetric capability to disentangle emission mechanisms that originate continuum emission (Synchrotron, Bremsstrahlung, Inverse Compton, neutral pions decay).

\section{Cosmic rays: de-excitation of nuclear lines}

On the other hand, The ``hadronic fingerprint'' can be read off the de-excitation lines of heavier nuclei: let us take the exemplary case of the Carbon line at 4.4 MeV from Cassiopeia A, considering the prediction from the hadronic fit ($p + A\longrightarrow \pi + A$) of GeV-TeV data \cite{9}. Let us take a proton power law $Q_p = Q_{0} p^{-2.3}$ with total energy $W_p = \int_{10 \rm{MeV}} Q_p p ~ dp = 4 \times 10^{49}$erg from \cite{9} and note that this would correspond to $\sim$2\% of the SNR kinetic energy (less than what is expected in the Ginzburg scenario \cite{10} \cite{11}).

\begin{figure}
\begin{center}

\psfig{figure=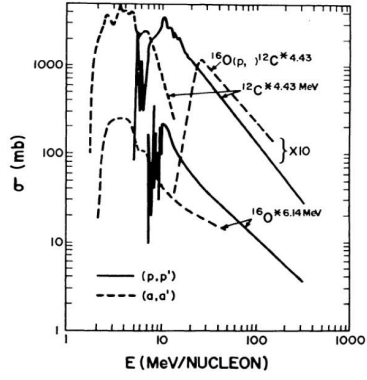,height=5.0cm,angle=0}
\vspace{-0.5cm}

\end{center}
\caption{\footnotesize
{Cross section for the $^{12}$C excitation and (quasi spontaneous) deexcitation line emission at 4.43 MeV, measured in a solar flare of August 1972 from \cite{14}.
}
}
\label{ram}
\end{figure}

According to \cite{12}, electrons with a similar slope produce the hard X-rays via synchrotron emission: the Inverse Compton emission is suppressed, owing to a strong magnetic field $B > 0.5 \rm{mG}$ (for leptonic GeV-TeV fits, 0.1 mG is needed). Let us mention that hadronic CRs at low energies could also be suppressed if their acceleration gains do not overcome Coulomb losses \cite{13} but we focus on the emission from the acceleration zone here.
Considering a heavy ion enriched plasma composition with $n_C =~10 \rm{cm}^{-3}$ at the reverse shock, the $^{12}$C excitation and (quasi spontaneous) de-excitation line emission at 4.43 MeV is given by:

$$ \frac{dN}{dE} (^{12}C \longrightarrow ^{12}C^{\ast}) = \int_{10 \rm{MeV}} Q_p (p) \frac{d \sigma (p)}{dp} n_C c ~ dp $$

where the cross section is shown in Fig. \ref{ram}.

Using d = 3.4 kpc for the distance to Cassiopeia A, this yields the line flux at 4.43 MeV:

$$ F_{4,43} = \frac{1}{4 \pi d^2} \frac{dN}{dE} \simeq 4 \times 10^{-8} \rm{cm}^{-2} \rm{s}^{-1} \rm{keV}^{-1}$$

Considering now a line width of E = 100 keV, we obtain the flux that we were searching for and let us note that such a flux is clearly detectable by GRM-GRIPS; as further check we can also compare it with the upper limit provided by COMPTEL of the flux in the range (3-10) MeV for the continuum \cite{15}:

$$ \Phi_{3-10} = 4 \times 10^{-6} \rm{cm}^{-2} \rm{s}^{-1} <   \Phi_{U.L.} = 1.4 \times 10^{-5} \rm{cm}^{-2} \rm{s}^{-1}$$

\begin{figure}
\begin{center}

\psfig{figure=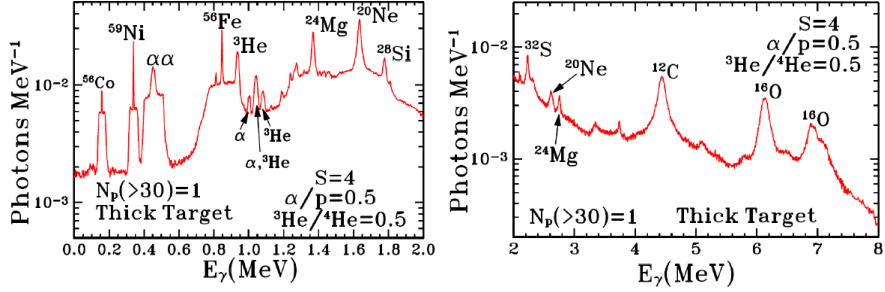,height=4.0cm,angle=0}
\vspace{-0.5cm}

\end{center}
\caption{\footnotesize
{Many other de-excitation lines could be visible in GRM-GRIPS energy range: above all $^{12}$C and $^{16}$O \cite{14}, Solar System abundances have been considered in computing this spectrum from \cite{14}.
}
}
\label{ram2}
\end{figure}

Figure \ref{ram2} depicts the predicted line spectrum around the $^{12}$C line according to \cite{14}.


To validate the chances for success of the introduced approach, we used the Monte-Carlo-code developed by Ramtay et al. to compute the whole nuclear de-excitation line spectrum for the specific case of Cassiopeia A. For a detailed description of the methods used in the code and a deeper insight into the different reaction types as well as the derivation of line production cross sections, we refer the reader to \cite{14} and \cite{koz}. 

\begin{figure}
\begin{center}

\psfig{figure=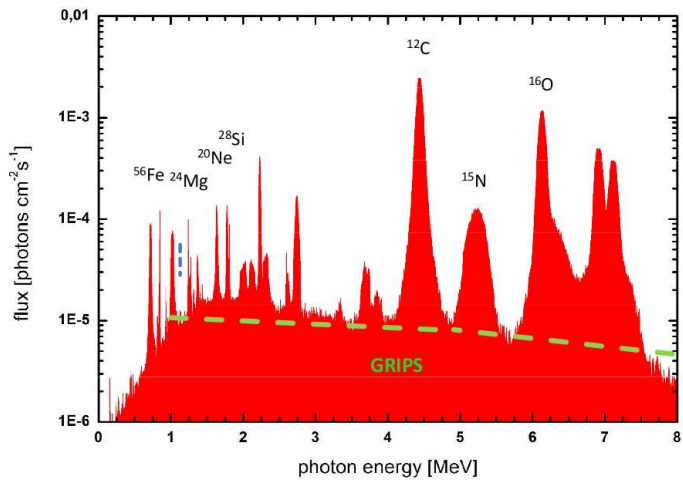,height=6.5cm,angle=0}
\vspace{-0.5cm}

\end{center}

\caption{\footnotesize
{ Spectrum for the specific case of Cassiopeia A abundances \cite{16}, in good agreement with the calculation above. The green dashed line represents GRIPS sensitivity in $\sim$ days of observations.
}
}
\label{red}
\end{figure}

The spectrum shown in Fig. \ref{red} has been obtained by extrapolating the high-energy proton spectrum inferred from Fermi and MAGIC data \cite{9} down to the MeV range. It is commonly argued that strong Coulomb losses would quench the particle spectra below GeV energies, but the steep spectra of non-thermal particles in solar flares may tell a different story. Furthermore kinetic models of particle acceleration in supernova remnants produce cosmic ray momenta spectra which are steeper in the low energy part than in the high energy part (e.g. \cite{v}). Although some non-thermal emission is associated with the forward shock, recent studies showed that electron acceleration to multi-TeV energies is likely to take place at the reverse shock in the supernova ejecta \cite{h}. Assuming also an acceleration scenario for cosmic rays at the reverse shock side, we adopt a chemical composition which is interstellar for the cosmic rays and Wolf-Rayet-like for the gas, consistent with X-ray data \cite{16}, visible in Tab. \ref{tab}. The resulting spectrum shows good agreement with the calculation above, showing that cosmic rays are able to produce a flux of nuclear de-excitation lines which would be clearly detectable by a future telescope mission in the MeV range. A unique feature arises due to C and O lines in the 4-6 MeV band, dominating the line flux from the Ne-Fe group in the 1-3 MeV band. From the gamma-ray measurements, also the spallation rate of heavy nuclei and thereby the effect on the abundances of Li, Be, and B could be inferred, providing a new assessment of their primordial abundances.




\begin{table}
\begin{center}
\begin{tabular}{|c|c|c|c|c|c|c|}
\hline
ratio & mean & rms\tabularnewline
\hline
\hline
O/Si & 1.69 & 1.37\tabularnewline
\hline
Ne/Si & 0.24 & 0.37\tabularnewline
\hline
Mg/Si & 0.16 & 0.15\tabularnewline
\hline
S/Si & 1.25 & 0.24\tabularnewline
\hline
Ar/Si & 1.38 & 0.48\tabularnewline
\hline
Ca/Si & 1.46 & 0.68\tabularnewline
\hline
FeL/Si & 0.19 & 0.65\tabularnewline
\hline
FeK/Si & 0.60 & 0.51\tabularnewline
\hline
Ni/Si & 1.67 & 5.52\tabularnewline
\hline
\end{tabular}
\end{center}
\caption{\footnotesize
{ Mean measured abundance ratios with rms scatter for Cassiopeia A ejecta \cite{16} used to compute Fig. 5 spectrum.
}
}
\label{tab}
\end{table}


\section{Conclusions}

The computed the nuclear de-excitation line spectrum concludes that with GRIPS we will be able to:

\begin{itemize}
\item {probe the hadronic acceleration in SNRs in independent ways (e.g. continuum spectrum, $^{12}$C and $^{16}$O lines);}
\item {study the elements abundances by studying the line ratios.}
\end{itemize}

We predicted the nuclear de-excitation line emission arising from interactions with the heavy elements in the supernova ejecta and the illustrative $^{12}$C example highlighted the importance of MeV gamma ray observations of the hadronic fingerprint of cosmic rays. We showed as well how these de-excitation lines could be observed in the MeV band with a future space mission such as GRIPS.

\end{document}